# Physiologic Blood Flow is Turbulent: Revisiting the Principles of Vascular Hemodynamics

Khalid M. Saqr[1*], Simon Tupin[1], Sherif Rashad[2,3], Toshiki Endo[3], Kuniyasu Niizuma[2,3,4], Teiji Tominaga[3], Makoto Ohta[1]

[1] Institute of Fluid Science, Tohoku University, Sendai, 980-8577, Sendai 980-8577, Miyagi, JAPAN.
[2] Department of Neurosurgical Engineering and Translational Neuroscience, Tohoku University Graduate School of Medicine, Sendai 980-8574, Miyagi, JAPAN.
[3] Department of Neurosurgery, Tohoku University Graduate School of Medicine, Sendai 980-8574, Miyagi, JAPAN.
[4] Department of Neurosurgical Engineering and Translational Neuroscience, Graduate School of Biomedical Engineering, Tohoku University, Sendai 980-8574, Miyagi, JAPAN.

*Correspondence to: k.saqr@tohoku.ac.jp, kh.saqr@gmail.com

## ABSTRACT

Contemporary paradigm of vascular hemodynamics considers normal blood flow to be pulsatile laminar flow. Transition to turbulence can cause diseases such as atherosclerosis or brain aneurysms. Recently, we demonstrated the existence of turbulence in experimental models of brain aneurysm; in the aneurysm sac as well as in the main artery. Thus, we were intrigued to explore if such a long-standing assumption of the laminarity of blood flow could be challenged. We have used methods and tools from chaos theory, hydrodynamic stability theory and turbulence physics to explore the existence of turbulence in normal vascular blood flow. We used Womersley's exact solution of the Navier-Stokes equation with the HaeMed© database of physiologic blood flow measurements, to offer reproducible evidence for our findings, as well as evidence from Doppler ultrasound measurements from healthy volunteers. The tools we used to investigate the properties of blood turbulence are well established in the fields of chaos theory, hydrodynamic stability and turbulence dynamics. We show, evidently, that blood flow is inherently chaotic and turbulent and not laminar. We propose a paradigm shift in the theory of vascular hemodynamics which requires rethinking the hemodynamic-biologic links governing physiologic and pathologic processes.

**INTRODUCTION**

The pioneering work of J.R. Womersley[1] and his coworkers[2,3] essentially laid the foundation of modern hemodynamics research. By developing a time-dependent one-dimensional exact solution of the Navier-Stokes equation, Womersley showed that blood flow in main arteries can be described by a Fourier decomposition of the cardiac harmonics[3,4]. This work has been later extended to account for wall elasticity[5,6] and non-Newtonian blood viscosity[7]. The Womersley flow model (*WFM*) became the founding principle upon which modern blood hemodynamics studies are based. Researchers have assumed, based on *WFM*, that the blood flow is essentially pulsatile laminar, and transition to turbulence (or presence of disturbed flow, which is a poorly defined hemodynamic term often used in medical hemodynamics research) shifts the blood hemodynamics leading to the initiation of vascular diseases such as brain aneurysms or atherosclerosis[8,9]. This disease association is due to the mechano-sensing properties of endothelial cells that make them especially responsive to various flow properties[10]. Therefore, the accurate identification of blood hemodynamics is an essential step in characterizing flow regimes that would govern processes in physiology and pathology[9]. We have previously shown how inaccurate assumptions regarding blood viscosity and the nature of wall shear stress (WSS) have impacted cerebral hemodynamics research, leading to widely varying and inaccurate results[9]. Moreover, we recently discovered the occurrence of non-Kolmogorov turbulence and inverse energy cascade[11] in brain aneurysm idealized model under physiologic flow using laser particle imaging velocimetry (PIV)[12]. Interestingly, in the same work we discovered that the flow in the parent artery itself was also turbulent, but not exhibiting this inverse energy cascade phenomena[12]. This prompted us to evaluate the nature of turbulence in actual blood hemodynamics, and whether the assumptions made regarding the *WFM* are actually correct. It is noteworthy to mention that transition to turbulence was detected in intracranial aneurysm using experimental[13] and computational[14-18] techniques.

In the present work, we tested WFM for three properties of turbulence: sensitive dependence on initial conditions (SDIC), hydrodynamic stability and kinetic energy cascade. To offer the maximum possible reproducibility evidence for our work, we relied on open-access HaeMed® database for obtaining the normal *WFM* of the carotid artery in healthy human subjects. The same tools can be easily used to demonstrate resemblance with other arteries. In this work, we report comparative results from Womersley exact solution and DUS measurements in the carotid artery, and in the supplementary materials we show exact solution results in thoracic aorta, aorta and iliac arteries to complement our hypothesis. First, the WFM is tested for chaos using Lyapunov exponent test. Then, the hydrodynamic stability theory is adapted to test the susceptibility of *WFM* to space-time hydrodynamic instability. Thirdly, we show that *WFM* intrinsically exhibits non-Kolmogorov kinetic energy cascade. Moreover, we performed carotid Doppler ultrasound measurements (DU) on healthy volunteers (authors of this study) and performed the same analysis to further confirm our findings. We here demonstrate that the physiologic blood flow is inherently turbulent and not laminar based on the *WFM* itself.

## MATERIALS AND METHODS

For the sake of reproducibility, we have worked with very simple tools including a well-documented exact solution methodology of the Womersley equation, open-source physiological flow database, and open-source chaos analysis code and well-established methods of computing hydrodynamic instability and turbulent energy cascade. This section describes the details of these methods to provide the essential steps to reproduce our results.

- **Exact Solution of the Womersley Equation**

The Womersley equation that defines the velocity profile of any physiological flow in arteries can be written as: $\tilde{u}(r,t) = \frac{ik_s r^2}{\mu \Omega^2}\left(1 - \frac{J_0(\zeta)}{J_0(\Lambda)}\right)e^{i\omega t}$ (1) where $k_s e^{i\omega t}$ is the oscillating pressure gradient, $r$ is the artery radius, $\mu$ is the viscosity of blood, $\Omega = r\sqrt{\frac{\rho \omega}{\mu}}$ is the Womersley number, $J_0$ is the Bessel function of zero order and first kind, $\Lambda = \Omega\left(\frac{i-1}{\sqrt{2}}\right)$ is the complex frequency parameter, and $\zeta(r) = \Lambda \frac{r}{R}$ is the complex variable. The derivation of equation (1) from Navier-Stokes equation can be found in the original paper by Womersley[4] or and in more instructive details in[19]. The exact solution of equation (1) as initial-boundary value problem has been extensively reported in literature, however, with no evidence of checking the existence of turbulence and chaos. In this work, the initial-boundary conditions were obtained from the open-access database Hae-Med© (http://haemod.uk/). This database provides blood flow waveforms for 11 major arteries. The present work establishes the investigation on the carotid artery (using the database as well as via doppler U/S measurements from volunteers) and provides the exact solution results for three more arteries in supplementary materials. The waveforms were created using physiologically realistic model based on human data. Hae-med© database has been extensively validated in literature and confirmed to represent physiologically relevant and valuable data. However, Fourier decomposition revealed that the waveforms are limited to 40 harmonics. No institutional approval was required for the DU measurements as per Tohoku University and Kohnan hospital (the location of the DU measurements) ethics guidelines.

- **Lyapunov exponent calculations**

For any time series dataset, the rate of separation of infinitesimally close orbital trajectories are characterized by Lyapunov exponent where the initial separation is $\delta X_o$ and the divergence rate is $|\delta X(t)| \approx e^{\lambda t}|\delta X_o|$ where $t$ is time and $\lambda$ is the Lyapunov exponent. We have used the open-source code provided by Wolf et al[20] and described in his 1985 famous paper[21] to build the phase-space and orbits of the time series datasets obtained from the exact solution of the Womersley equation and DU measurements.

- **Evaluation of global hydrodynamic stability in pulsatile flow**

The onset of transition in physiological flow requires hydrodynamic instability to commence. Such instability is caused by a sustained disturbance in the flow field, where its energy is expressed as:

$$\int_0^\infty E_V dV = \frac{1}{2}\overline{\tilde{v}^2_{(x_i,t)}} \qquad (2)$$

where $\tilde{v}(x_i, t) \notin u(x_i, t) = u_o(x_i) + \sum_{i=1}^{n} a_i \cos(i\omega t) + b_i \sin(i\omega t)$ where $u(x_i, t)$ is the blood velocity waveform with $a_i$ and $b_i$ as the Fourier coefficients of the Womersley waveform, and $u_o(x_i)$ is the steady component of the flow which is characterized by Hagen-Poiseuille parabolic velocity profile. The sustenance of a disturbance must be global (i.e. in space and time) for transition to take place.

In order for the flow to become globally instable, the time rate of change of $E_V$ must be positive, hence the disturbance can possibly prevail over the viscous resistance of the fluid. The Reynolds-Orr equation [22] for disturbance energy reads:

$$\frac{dE_V}{dt} = -\int_V u \cdot (S_{ij} u) dV - \frac{2}{Re} \int_V s_{ij} : s_{ij} dV \qquad (3)$$

where $t$ is time, $S_{ij}$ and $s_{ij}$ are the strain rates of basic and altered (i.e. disturbed) flows, respectively, $Re$ is the Reynolds number and $V$ is the flow volume bounded by a wall $\ell$. The first term on the right-hand side (RHS) of (2) describes the exchange of energy with the basic flow, while the second term describes the energy dissipation due to viscosity. If the former is higher than the latter (i.e. the disturbance energy does not decay with time), the flow will be globally instable. Thus, the criteria for global stability can be written in terms of critical Reynolds number as:

$$\frac{1}{Re_{cr,m}} = \max_{v(x,t)=0} \frac{-\int_V u\cdot(S_{ij}u)dV}{2\int_V s_{ij}:s_{ij}dV} \text{ when } \frac{dE_V}{dt} = 0 \tag{4}$$

In equation (4), the critical Reynolds number $Re_{cr,m}$ defines the global and monotonic stability criterion for viscous fluid flow subjected to a disturbance of the velocity $u(x,t)$. Hence, if $Re > Re_{cr,m}$ the flow becomes monotonically instable, in other words the disturbance will not decay for $t \to \infty$ and transition will take place. Hence, asymptotic instability occurs when $\lim_{t\to\infty} \frac{E_V(t)}{E_V(0)} \neq 0$. In pulsatile multiharmonic flow, such condition takes the form $\lim_{t\to\infty} \frac{E_V(n)}{E_V(n-1)} \neq 0$ where $n$ is the index of harmonic in the Womersley velocity waveform described in equation (2). This condition has been evaluated in space and frequency domain using equation (2) and the trapezoidal method for numerical differentiation. Following Thomas[23,24] and Hopf[25], Serrin[26] proved that in order to have monotonic instability in any viscous flow, the disturbance energy must satisfy the following condition:

$$E_V > E_{VO}\, e^{\frac{t}{v}\left(U_{cr,m}^2 - Re_{cr,m}\frac{v^2}{d^2}\right)} \tag{5}$$

where $E_{VO}$ is the initial disturbance kinetic energy, $U_{mc}$ is the velocity of the basic flow during a one cycle of the pulsatile flow. Serrin's second theorem proved that $Re_{cr,m} = 32.6$ for viscous incompressible wall bounded flows. Reynolds number in main arteries is larger than the latter value. Hence, blood flow, by definition, is monotonically instable.

- **Calculating Energy Cascade in Frequency Domain**

The local kinetic energy in frequency domain was calculated as[27]: $E_i(f) = \frac{|\mathcal{F}\{\tilde{u}_i(t)\}|^2}{L\cdot f}$ where $f$ is the frequency, $\mathcal{F}$ is the fast Fourier transform and L the length of $\tilde{u}_i$ matrix. It was not possible to use Taylor's hypothesis to calculate the kinetic energy in wavenumber domain since $\frac{\tilde{u}}{U} > 1$. figure 1 shows the waveform profile to show that the conditions for using Taylor's hypothesis do not apply.

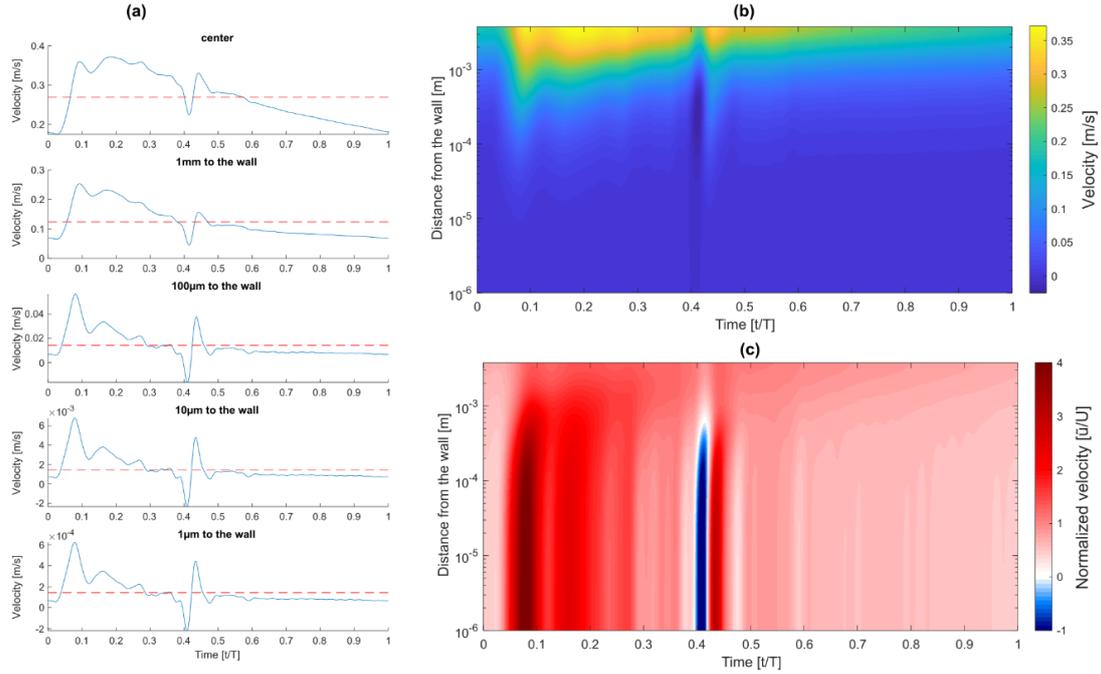

**Figure 1: The advection of oscillating harmonics in carotid waveform cannot be correlated with the mean flow velocity as long as $\frac{\tilde{u}}{U} > 1$**, hence, it is not possible to obtain the energy field in wavenumber domain. (a) The carotid waveform at varying distance from the wall is plotted against dimensionless time and the red dashed line depicts the mean flow velocity. The flow near the wall is more complex than such away from the wall. (b) Colourmap of the velocity field of one cycle plotted against distance from the wall. (c) Colourmap of $\frac{\tilde{u}}{U}$ plotted against distance from the wall.

## RESULTS

Herein, we present the data obtained from carotid artery analysis. Results from the exact solution of two more vessels are presented in the online supplementary data along with additional results explanation. We hypothesize that the intrinsic multiharmonic flow waveforms are the properties driving blood flow towards chaotic turbulent state. As much as it is important to characterize the differences between vessels in such regard, the authors believe that it could be cumbersome to conduct such characterization in this article. We invite the community to reproduce the following results using other sets of data for different arteries to further investigate blood turbulence.

- **Vascular Blood Flow is Chaotic**

First, we borrowed tools from the chaos theory to test for the SDIC[21], and to demonstrate that blood flow, expressed by both exact solution and *in vivo* measurements, is not periodic. We have obtained Lyapunov exponents, as a test of SDIC associated with turbulence[28], from the time series of blood velocity obtained from the Womersley equation and from DU measurements. Figure 2-a shows the trace of Lyapunov exponents in orbital space for the carotid artery. Similar Lyapunov exponents obtained from *in vivo* DU measurements are shown in figures (2-b) and (2-c) for two healthy volunteers (two of the authors). Positive Lyapunov exponents, indicating SDIC, has been previously reported in few studies investigating pulsatile flow in an artificial heart[29]. However, this is the first evidence on the existence of SDIC in Womersley equation. A more elaborative discussion of relevant researches on SDIC of blood flow is given in the supplementary materials.

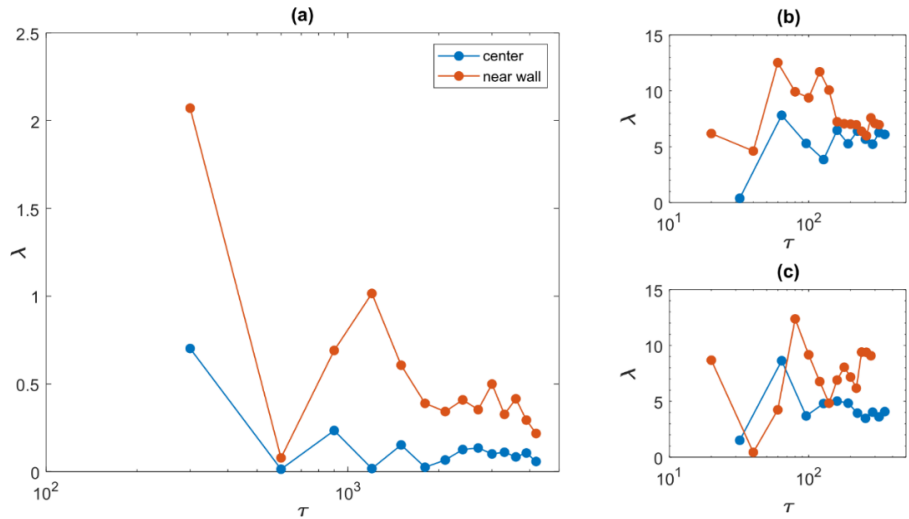

**Figure 2. Positive Lyapunov exponents indicate SIDC in both exact solution of the Womersley equation and DU in-vivo measurements.** (a) The orbital ($\tau$) traces of Lyapunov exponent ($\lambda$) from exact solution at the center and 1μm away from the wall of the Carotid artery. (b) and (c) Lyapunov exponents in the center and 1/3 D away from the wall of the carotid artery, respectively, in two healthy volunteers.

- **Vascular Blood Flow is Globally Unstable**

The theory of hydrodynamic stability has been developed to evaluate the stability of steady flow under finite space-time perturbation[30,31]. Some works have evaluated the stability of monoharmonic (i.e. sinusoidal) flow[32-36]. However, there are no universal criteria that can generally describe the stability of multiharmonic pulsatile flow. We have extended the available criteria, as explained in supplementary materials, to evaluate Womersley flow. In Womersley flow, the velocity field is given as:

$$u(x_i, t) = u_o(x_i) + \sum_{i=1}^{n} a_i \sin(i\omega t) + b_i \cos(i\omega t) \qquad (1)$$

where $n$ is the total number of harmonics in a specific waveform. Here, we propose that a given Womersley flow is globally unstable if $\lim_{t \to \infty} \frac{E_V(n)}{E_V(n-1)} \neq 0$. Figure 3 shows colourmap of such condition as obtained both from exact solution and *in vivo* DU measurements of the carotid artery. It is clear that the condition is achieved in both datasets confirming that blood flow is globally unstable. In figure 4, the convective acceleration field obtained from the exact solution is depicted. The maximal and minimal acceleration values are proximal in time domain, while the remaining time domain of the acceleration field is dominated by quasi-steady flow. This supports the assumption made to evaluate the global instability, bearing in mind that $\int_0^\infty E_V dV = \frac{1}{2}\overline{\tilde{v}_{(x_\iota,t)}^2}$ is primarily affected by convective acceleration $\vec{v} \cdot \nabla \vec{v}$.

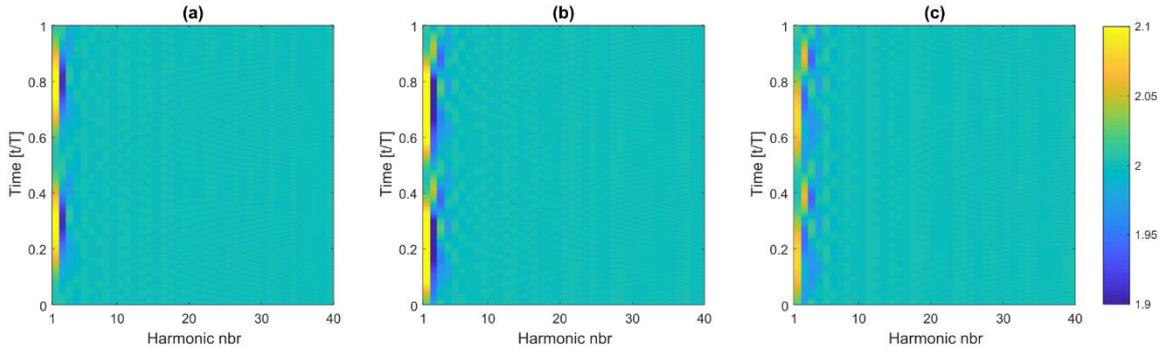

**Figure 3. Global hydrodynamic instability condition achieved in blood flow waveforms obtained from (a) exact solution of Womersley equation and (b,c) in-vivo DU measurements.** Harmonic number represents the number of harmonics in one cardiac cycle in both datasets and the colourmap represents the values of $\lim_{t \to \infty} \frac{E_V(n)}{E_V(n-1)} \neq 0$ corresponding to 40 harmonics which constitute the pulsatile waveform.

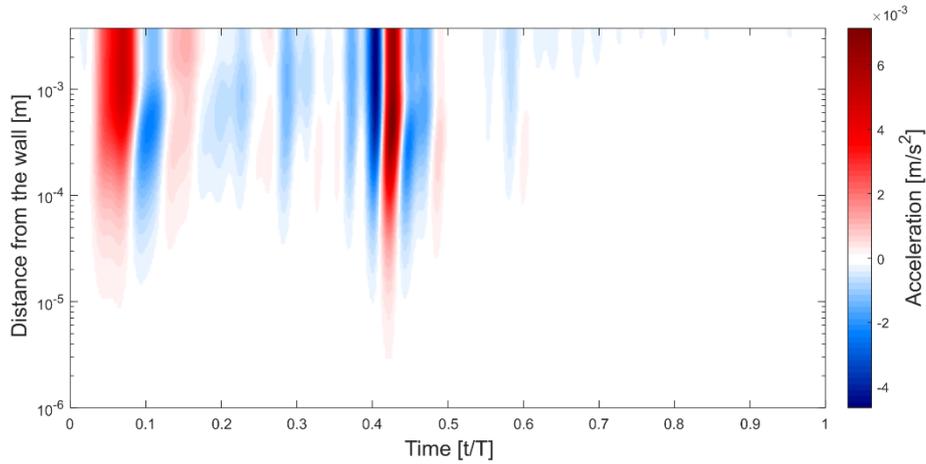

**Figure 4: Acceleration field obtained from the Womersley exact solution.** The field is dominated by quasi-steady flow $v \cdot \nabla v \approx 0$ while peaks of minimal and maximal acceleration reciprocate at distance corresponding to $5 \times 10^{-5}$ from the wall. Acceleration alters the flow energy significantly driving it towards global instability.

- **Normal Vascular Blood Flow Exhibits Kinetic Energy Cascade**

One of the most profound properties of turbulence is the cascade of kinetic energy. Kinetic energy is transferred from larger to smaller vortex structures and ends up irreversibly dissipating to the surrounding media in the form of heat. Figure 5 shows kinetic energy cascade in carotid blood flow as depicted from exact solution and *in vivo* DU measurements. It can be seen that the rate of such cascade does not match any of the Kolmogorov scales reported in literature. Turbulence, in such case, is of non-Kolmogorov regime. Such uncharacteristic turbulence regime has been previously observed atmospheric turbulence[37,38], however, it is the first time to be reported in the human circulatory system. Until the present moment, the main mechanism linking hemodynamics to biological and pathological processes in arteries is the wall shear stress[8]. The existence of kinetic energy cascade in blood flow should develop this mechanism to include other mechanical and kinetic factors that could contribute to mechanosensory and mechanotransduction of endothelial cells.

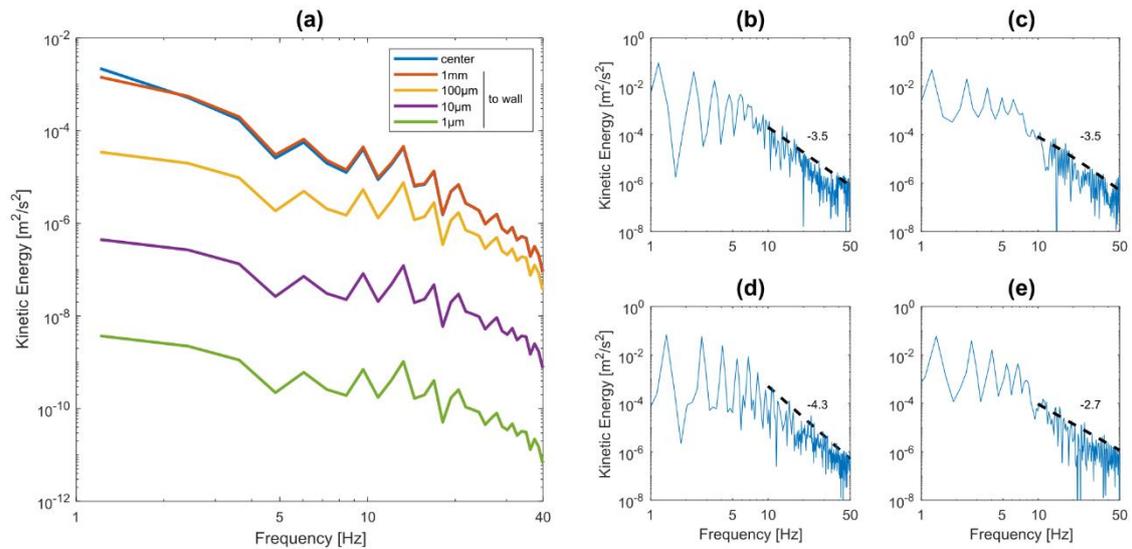

**Figure 5. Kinetic energy cascade detects non-Kolmogorov kinetic energy transfer at frequencies higher than dominant frequency.** In (a), the kinetic energy cascade obtained from the exact solution of Womersley flow of carotid artery are plotted at varying locations from the wall. Similar cascades are displayed for the carotid blood flow of two healthy volunteers at the artery center (b and d) and at $\frac{1}{3}D$ away from the wall (c and e).

## DISCUSSION

Almost a decade earlier to the publication of WFM in *the Journal of Physiology*, the work of Kolmogorov on turbulence was published in Russian and translated to English. Kolmogorov explored the statistical isotropy and homogeneity of turbulence in fluids and deduced the famous scaling laws to describe such properties. It was thought that since *WFM* describes quasi-periodic flow in cylindrical conduits (i.e. vessels) with Reynolds number below 1000, it would not fit into the Kolmogorov theory of statistically isotropic and homogenous turbulent flow. Hence, the classification of Womersley flow was established to be laminar pulsatile flow in normal human arteries[39,40]. The deviation from the so called laminar flow conditions is believed to be linked to atherogenesis and aneurysm formation through various mechanisms associated with endothelial cells inflammatory reactions and dysfunction[41-45]. These assumptions have dominated the vascular biology research, and most of the body of literature regarding endothelial cells responses to hemodynamics have started from such a view of the *WFM* [8,46]

Turbulent flow has four characteristics that keeps it as the last standing mystery in classical mechanics[47]. These properties are chaos, instability, rotationality and kinetic energy cascade[48]. Turbulent flow is sensitive to initial conditions, which make it subjected to the properties of chaos theory[49]. A turbulent field is instable where any finite perturbation in space and time propagates. The vorticity field is always non-zero in turbulent flows, and the kinetic energy transfers from large to small vortices. The kinetic energy cascade is best described in frequency and wavelength domains. In our analysis of the Womersley equation and *in vivo* Doppler measurements, we performed Lyapunov exponent analysis, which essentially informs us whether the data or equation are inherently chaotic or not. Womersley equation had a positive Lyapunov exponent, meaning that it has properties of chaos under multiharmonic pulsatile flow that is the blood flow. This, by definition, negates the assumption that the *WFM* describes a pulsatile laminar blood flow regime. In fact, we discovered that the blood flow is globally turbulent under physiologic conditions. Thus, a transition of turbulence *per se* is not the causative factor behind disease initiation, rather a change in turbulence characteristics due to complex geometry or harmonics would be the culprit. This of course adds a layer of complexity to vascular hemodynamics and biology.

When one thinks of laminar pulsatile flow, the only available flow field variable relevant to physiological vessel changes is wall shear stress, which is well studied and characterized especially in *in vitro* endothelial flow exposure experimentation[10]. On the other hand, when turbulent flow is considered, numerous variables could affect the vessel on multiple scales. For instance, kinetic energy, represented by the energy cascade, transfers from large coherent vortices to small isotropic homogenous vortices before dissipating in viscosity scales as heat. We have evaluated the length scales and energy scales of such vortices and they fall well within the limits of cellular mechano-sensing[50]. This begets an important question, how do different energy cascades, for example a direct versus an inverse cascade, impact endothelial cells? and more importantly, how can we fine tune our experimental setups to measure and reproduce these conditions? Moreover, how would these new findings impact vascular hemodynamics based therapeutic such as shear driven drug deliver, a possible drug delivery method to areas of altered shear stress[51], or intravascular devices[52]? All of these questions, indeed, are interesting prospects of the presented results.

In conclusion, we propose a paradigm shift for vascular hemodynamics. That is, the blood flow is inherently turbulent and not laminar, and changes in turbulence kinetic energy or other properties of turbulence could be the driving factors behind the hemodynamics-biology links. This paradigm shift must motivate us to update our thinking regarding hemodynamic drivers of endothelial and vascular processes, given the inherent complexities and chaos associated with turbulence. Moreover, more hemodynamic research, with accurate and rigorous methodologies, should aim at further characterizing the interesting features of turbulent blood flow in physiology and pathology.

**Acknowledgments:** The authors acknowledge the help of Dr. Alan Wolf by proving the code for Lyapunov exponents to public domain.

**Funding:** Collaborative Research Grant (Project Code: J19I001) from the Institute of Fluid Science, Tohoku University and JSPS KAKENHI [grant number 18K18356], JAPAN.

**Author contributions: Saqr:** Conception, methodology and writing. **Tupin:** Formal analysis, investigation, software and visualization. **Rashad:** Project administration and writing. **Endo:** Ultrasound measurements and validation. **Niizuma:** Validation, review and editing. **Tominaga:** Review and editing. **Ohta:** Funding acquisition, resources, review and editing.

**Competing interests:** The authors report no conflict of interest regarding this work nor there are any ethical or financial adherences.

**Data and materials availability:** Raw data is available upon request.

# Physiologic Blood Flow is Turbulent: Revisiting the Principles of Vascular Hemodynamics

## SUPPLEMENTARY MATERIALS

Khalid M. Saqr[1*], Simon Tupin[1], Sherif Rashad[2,3], Toshiki Endo[3], Kuniyasu Niizuma[2,3,4], Teiji Tominaga[3], Makoto Ohta[1]

[1] Institute of Fluid Science, Tohoku University, Sendai, 980-8577, Sendai 980-8577, Miyagi, JAPAN.
[2] Department of Neurosurgical Engineering and Translational Neuroscience, Tohoku University Graduate School of Medicine, Sendai 980-8574, Miyagi, JAPAN.
[3] Department of Neurosurgery, Tohoku University Graduate School of Medicine, Sendai 980-8574, Miyagi, JAPAN.
[4] Department of Neurosurgical Engineering and Translational Neuroscience, Graduate School of Biomedical Engineering, Tohoku University, Sendai 980-8574, Miyagi, JAPAN.

*Correspondence to: k.saqr@tohoku.ac.jp, kh.saqr@gmail.com

## 1. Comments on the Lyapunov exponents of blood flow waveforms

Sensitive dependence on initial conditions (SDIC) in blood flow has been the focal interest of few studies along the past three decades. Yip et al[1] found positive Lyapunov exponents in the renal artery of spontaneously hypertensive rats (SHR). Their results indicated the presence of a low-dimension strange attractor. While studying artificial heart systems, Yambe et al[2] discovered that blood flow waveforms have positive Lyapunov exponents and investigated the phase portraits of the dynamical system composed by the flow waveform. They demonstrated the existence of lower dimensional chaotic dynamics in blood flow. The works of Yin et al and Yambe et al used multiharmonic flow functions of the same form used in the present work (equation 1), however, their estimation of Lyapunov exponent was space-independent. When polynomial functions are used to describe blood flow waveforms, such as in the works of Bračič and Stefanovska[3,4], Lyapunov exponents come in pairs showing the periodic nature of these functions. Also, the level of details of the harmonic function is crucial to describe the actual physics of the flow. Here, he have used 40 harmonics to compose the solution of equation (1). We argue here that positive Lyapunov exponents, as an indicator of SDIC, are intrinsic properties of the Womersley equation given that sufficient number of harmonics is used to describe blood flow accurately. This is inductively shown by the *in vivo* data. Visee et al[5] used consecutive transcranial Doppler (TCD) to investigate the existence of chaos in patients with occlusive cerebrovascular disease. They had evidently showed that blood flow in healthy conditions exhibited positive Lyapunov exponents, while in impaired cerebral circulation it was found to be of periodic nature with near-zero Lyapunov exponents. These findings have been also supported by the works of Ozturk[6] and Ozturk and Arslan[7] where TCD signals showed positive Lyapunov exponents in the cerebral circulation.

## 2. Comments on the hydrodynamic stability of Womersley equation

In fluid mechanics, it is generally believed that hydrodynamic instability is always associated with turbulence[8]. However, most of the work done to understand the correlation between the two phenomena in pipe flow was mostly limited to steady flows subjected to finite perturbations in space and time[9]. The stability of pulsatile flows have been studied in some works[10-13], however, there is no consensual methodology among fluid dynamicists for investigating their stability. Therefore, to study the global instability of blood flow both from Womersley equation and *in vivo* measurements, we had to adopt the main principles of the hydrodynamic stability theory and develop intuitive representation of its criteria for global instability. The concept underlying this approach is fairly simple. Serrin[14] argued that for a viscous flow to be considered stable *"the energy of any disturbance tends to zero as t increases"*. Hence, we assumed that each harmonic component of the flow waveform $(n)$ could be viewed as a space-time perturbation to its predecessor in in frequency domain $(n-1)$. In supplementary figure 1, the stability condition derived from this principle is plotted for four arteries based on the exact solution of Womersley equation using boundary conditions from Hae-Med® database. We have used 40 harmonics to represent the solution of Womersley equation. The reasoning behind this is based on a mass transfer analysis. We have analyzed the contribution of each harmonic in mass transfer by computing the mass flow rate $\dot{m} = A \times u(x_i, t)$ with a loop that changes the number of harmonics $(n)$ from 1 to 1000. We have found that $\lim_{n \to \infty} \frac{\dot{m}_{n+1} - \dot{m}_n}{\dot{m}_n} = 0$. Supplementary figure 2 shows this difference for the first 40 harmonics. Based on dimensional analysis, we assume that $\lim_{n \to 40} \frac{\dot{m}_{n+1} - \dot{m}_n}{\dot{m}_n} = 10^{-8}$ is an appropriate cut threshold.

## 3. Comments on the kinetic energy cascade of Womersley exact solution

The cascade of kinetic energy is a characteristic feature of turbulent flows and it contributes to the very definition of turbulence[15,16]. Steady, periodic and quasi-periodic laminar flows tend to conserve their kinetic energy due to the absence of the vortex formation and breakup phenomena. The classical Kolmogorov-Obukhov theory of turbulence describes homogenous isotropic turbulence in which energy cascade is subjected to scaling laws. This is not the case in hand. Supplementary figures 4 and 5 show the kinetic energy cascade, in frequency domain, for four arteries at two different radial locations based on the Womersley exact solution.

It is clear from both figures that the scaling power of kinetic energy cascade is affected by spatial location (i.e. radial distance from the wall). Also, it is clear that the values of scale power is far from the values commonly accepted in the classical theory of turbulence[17,18]. This suggests that the turbulence observed here is non-Kolmogorov turbulence. Our group has recently shown *in vitro* that blood flow in ideal cerebral aneurysm model undergoes non-Kolmogorov turbulence[19]. The spikes in the energy cascade might indicate relevance to hydrodynamic resonance resulting from self-sustained oscillations arising from the multiharmonic pulsatile flow[20]. It should also noteworthy to mention the resemblance between energy cascade scaling power in one and five pulses, and in spatial location. Moreover, non-Kolmogorov turbulence was detected in carotid artery stenosis by Lancellotti et al using Large Eddy Simulation[21] and Mancini et al using Direct Numerical Simulation (DNS) and LES[22]. Here we show that it exists both *in doctrina* and *in vivo*.

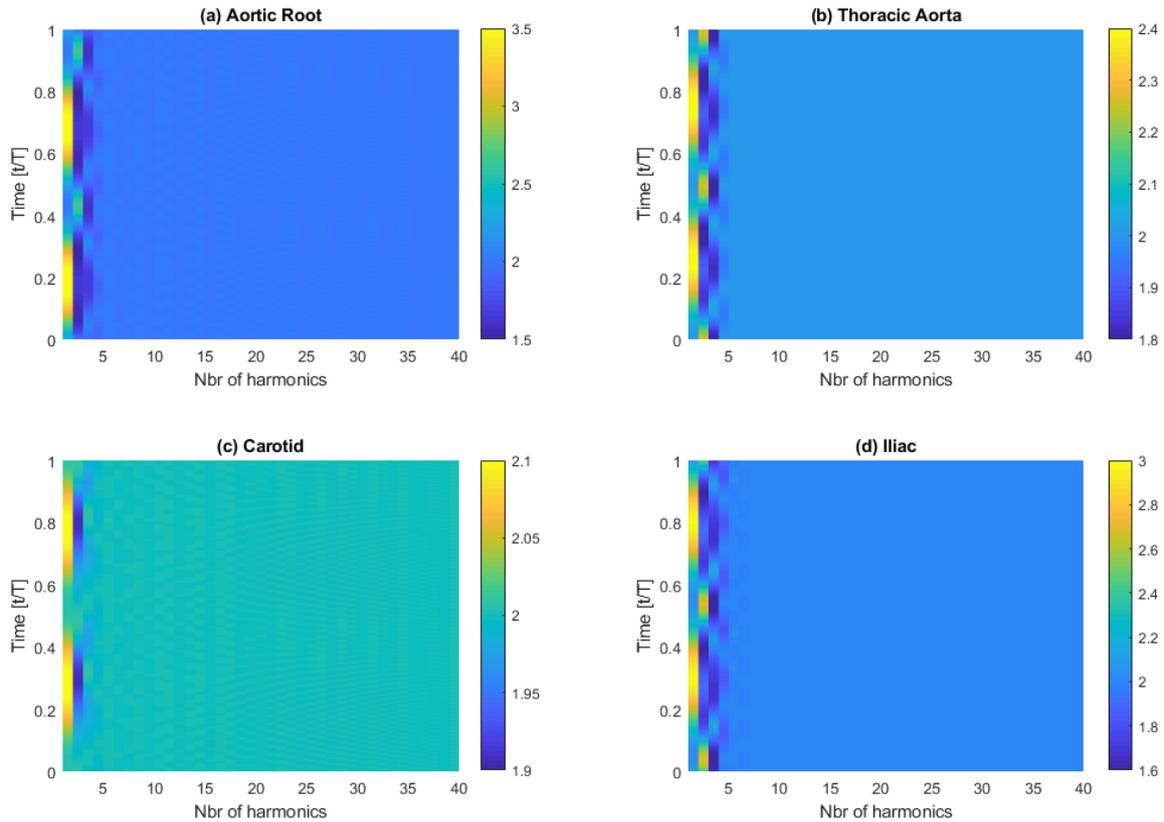

**Supplementary figure 1.** Color maps of $\lim_{t \to \infty} \frac{E_V(n)}{E_V(n-1)} \neq 0$ in from the exact solution of Womersley flow equation in (a) aortic root (b) thoracic aorta (c) carotid artery and (d) iliac artery.

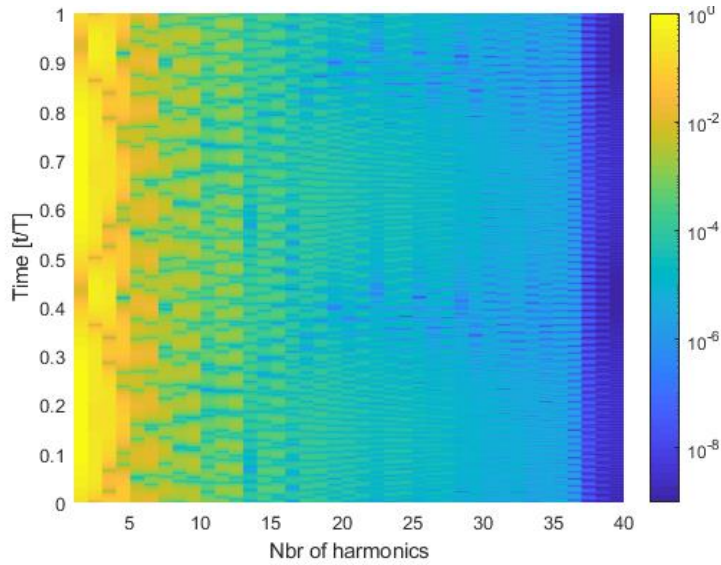

**Supplementary figure 2.** Color map of $\lim_{n\to\infty} \frac{\dot{m}_{n+1}-\dot{m}_n}{\dot{m}_n}$ for $n \leq 40$ on x-axis and non-dimensional time on y-axis.

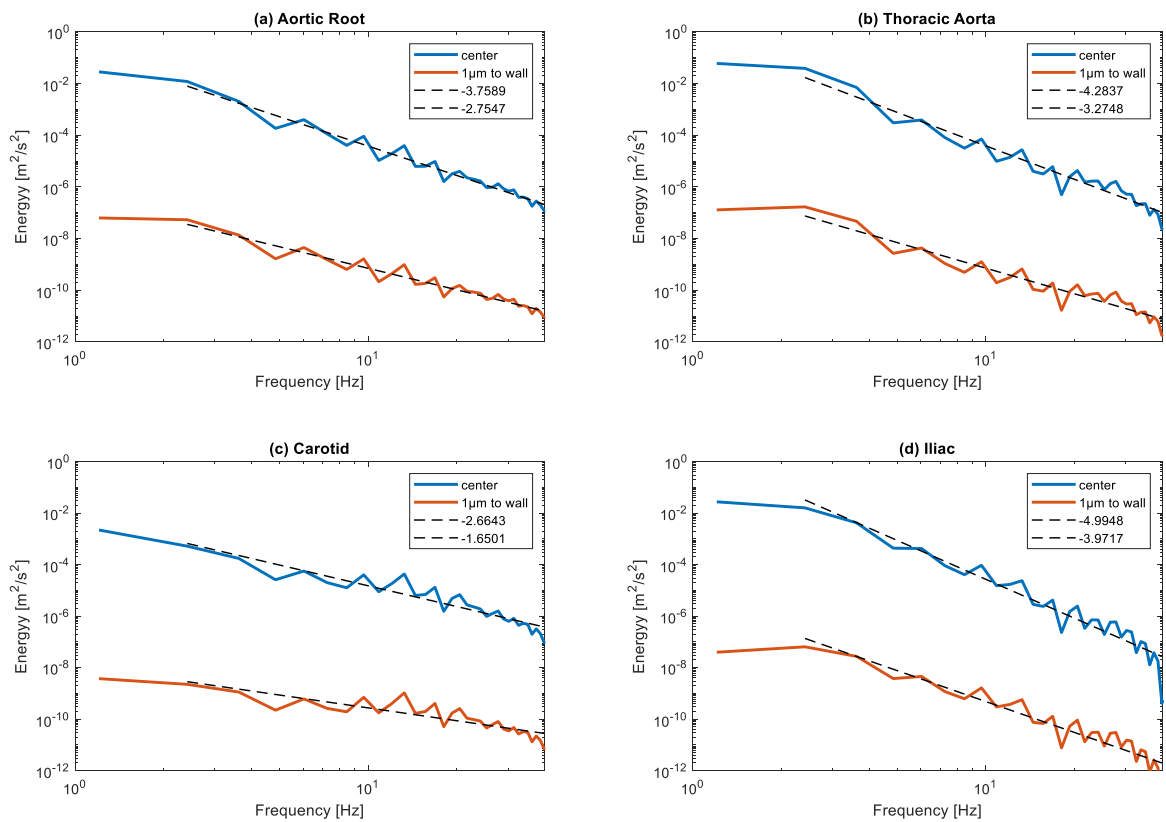

**Supplementary figure 3.** Kinetic energy cascade from one pulse of the Womersley exact solution based on the boundary conditions representing (a) aortic root (b) thoracic aorta (c) carotid artery and (d) iliac artery. Solid lines represent the energy cascade at the center (blue) and $1\mu m$ from the wall (red) and dashed lines represent the log fitting to show the slope of the cascade expressed in the inset legend in decimal fraction.

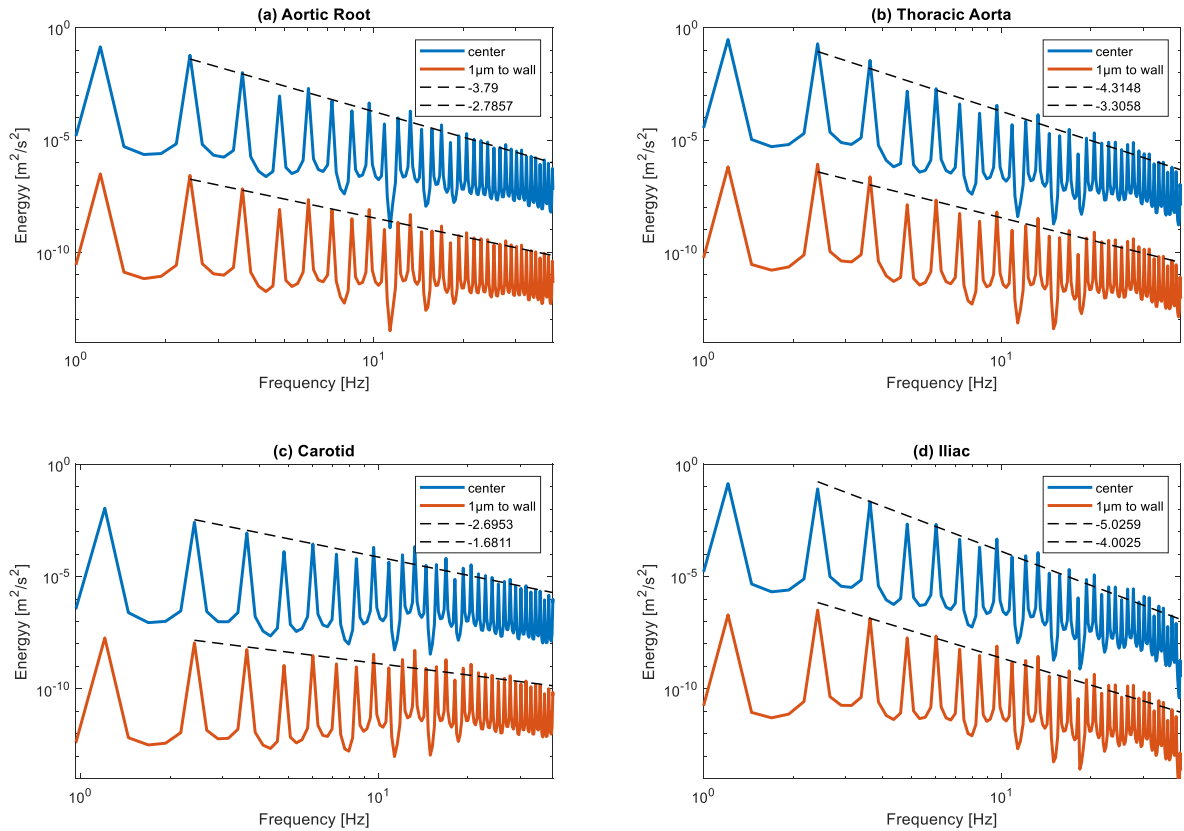

**Supplementary figure 4.** Kinetic energy cascade from five pulses of the Womersley exact solution based on the boundary conditions representing (a) aortic root (b) thoracic aorta (c) carotid artery and (d) iliac artery. Solid lines represent the energy cascade at the center (blue) and $1\mu m$ from the wall (red) and dashed lines represent the log fitting to show the slope of the cascade expressed in the inset legend in decimal fraction.

**SUPPLEMENTARY REFERENCES**